# Assessing Text Mining and Technical Analyses on Forecasting Financial Time Series

**Ali Lashgari***

**Department of Economics**
**Kansas State University, Manhattan, KS**
**January 2023**

**ABSTRACT**

Forecasting financial time series (FTS) is an essential field in finance and economics that anticipates market movements in financial markets. This paper investigates the accuracy of text mining and technical analyses in forecasting financial time series. It focuses on the S&P500 stock market index during the pandemic, which tracks the performance of the largest publicly traded companies in the US. The study compares two methods of forecasting the future price of the S&P500: text mining, which uses NLP techniques to extract meaningful insights from financial news, and technical analysis, which uses historical price and volume data to make predictions. The study examines the advantages and limitations of both methods and analyze their performance in predicting the S&P500. The FinBERT model outperforms other models in terms of S&P500 price pre- diction, as evidenced by its lower RMSE value, and has the potential to revolutionize financial analysis and prediction using financial news data.

---

*Thanks to Professor Yoonjin Lee

## INTRODUCTION

Forecasting is a useful tool in a variety of academic fields, from geography to business. Forecasting can help geographers predict changes in weather patterns, natural disasters, and land use, which allows for better urban planning and dis- aster preparedness [1]. It could be used to predict energy consumption patterns and demand in urban areas, which could inform the design of an appropriate lighting plan or transportation energy [2-4].

Forecasting is critical in mathematics and physics for understanding and predicting complex phenomena. Forecasting is used in physics to model and predict the behavior of physical systems like quantum mechanics and fluid dynamics. By tapering the end of photonic bandgap fibers, researchers created gas-filled photonic microcells with high transmission efficiency and moderate line center accuracy. The cells could be linked, albeit with a slight decrease in efficiency. This advancement has the potential to be used to monitor and predict changes in the performance of PMCs over time, potentially improving the efficiency and accuracy of laser micro/nano-machining processes [5-8].

Physicians can utilize forecasting to improve the prevention and management of PCOS and T2DM. Forecasting can aid in predicting an individual's risk of developing these conditions based on lifestyle, genetic, and other relevant factors. This information can then be used by physicians to create personalized prevention and management plans for their patients. For example, if a person is at a higher risk of developing T2DM, the doctor may recommend specific lifestyle changes and/or medications to prevent or delay the disease's onset. Similarly, if a patient has PCOS, forecasting can be used to predict how the patient will respond to various treatments and help the doctor tailor the treatment plan accordingly [9-10].

Forecasting using computational methods is essential in engineering as it enables better decision-making and efficient utilization of resources. Forecasting in the financial time series (FTS) is a crucial area in finance and economics that predicts potential risks, market trends, and investment timing in financial markets. This subject has been widely used for decades due to the uncertain and noisy nature of the financial environment. In order to analyze and predict financial market behavior, fundamental and technical analysis are commonly used.

Fundamental and technical analysis were widely used to analyze and forecast financial market behavior prior to the introduction of natural language processing (NLP) models. Fundamental analysis entails examining financial metrics and indicators to determine a company's financial health and growth potential, whereas technical analysis employs historical price and volume data to identify trends and forecast future market behavior. Both methods have benefits and drawbacks and can be used together or separately, depending on an investor's goals and investment style. Predicting financial market behavior, on the other hand, is difficult and is influenced by a number of factors such as economic indicators, geopolitical events, and investor sentiment. As a result, when making investment decisions, investors should consider multiple sources of information and analysis(Figure1).

The rapid spread of big data and artificial intelligence in recent years has resulted in an increasing amount of financial data and a more complex correlation mode between data, making forecasting financial time series data more difficult. Nonparametric methods that are not bound by statistical assumptions are required, and as a result, machine learning algorithms have become a focus for experts. Text mining, which employs NLP techniques to analyze massive amounts of textual data, has evolved into a powerful tool for time series fore- casting, serving as a supplement to traditional statistical methods. The primary benefit of text mining analysis is its ability to quickly analyze large amounts of textual data and identify trends and patterns in the data, which can be used to identify relationships between sentiment expressed in textual data and stock market movements [11-18].

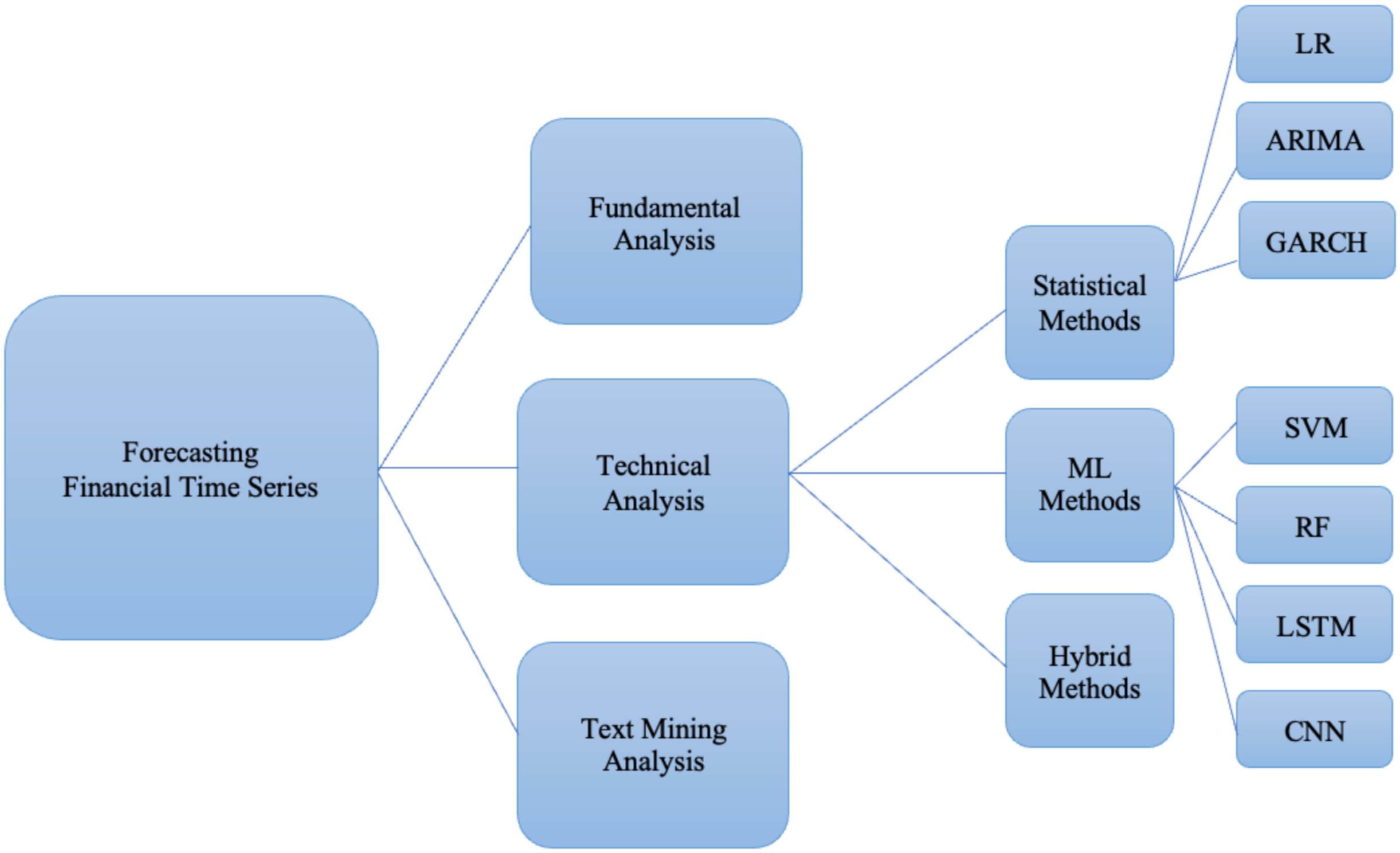

Figure 1: FTS

This paper investigates the accuracy of different technical methods and text mining for predicting S&P500 using financial news and historical price data. The proposed text mining method shows some improvement when compared with similar previous studies, and the results can help investors make better decisions when faced with unpredictable events.

By comparing these two different methodologies, we can determine which approach is more reliable in real-world problems - historical prices or news. This comparison can help investors decide which method to use when predicting future values in unpredictable events. Ultimately, the question of whether price or news is a better predictor of the future is a complex one, and the answer likely depends on the specific situation. However, this paper's findings can provide valuable insights into the strengths and weaknesses of these different approaches and help investors make more informed decisions.

The following is how the paper is structured. The related work is described in the section that follows. The proposed methodology, evaluation metric, and data set are all covered in Section 3. The experimental setup and results are shown in Section 4. Section 5 finally exposes our conclusions and future work.

## LITERATRE REVIEW

Word embedding is a technique in natural language processing (NLP) for representing words as numerical vectors. The goal of word embedding is to capture the meaning of words in a way that can be used as input to machine learning models.

Each word in a text corpus is represented as a high-dimensional vector, where each dimension corresponds to a particular semantic or syntactic property of the word. The values of the vector are learned from the training data, and the vectors are designed to capture the relationships between words in the corpus. For example, words that are semantically similar are expected to have similar vectors, while words that are semantically dissimilar are expected to have dis- similar vectors.

There are several types of word embedding techniques that are commonly used in natural language processing (NLP). Some of the most popular ones include:

Word2Vec: Published in 2013 by Tomas Mikolov et al. [19]
GloVe: Published in 2014 by Jeffrey Pennington et al. [20]
FastText: Published in 2016 by Piotr Bojanowski et al. [21]
ELMo: Published in 2018 by Matthew Peters et al. [22]
BERT: Published in 2018 by Jacob Devlin et al. [23]
GPT: Published in 2018 by Radford et al. [24]

The study by Atkins, Niranjan, and Gerding (2018) provides valuable in- sights into the predictive power of financial news compared to the close price of assets or indexes. The authors use machine learning models to analyze news feeds and predict the direction of asset price movement and asset volatility movement. Their findings reveal that news-derived information is a better predictor of market volatility than the close price of an asset or index, with an average directional prediction accuracy of 56 [12].

The study by Souma, Vodenska, and Aoyama (2019) explores the use of deep learning for sentiment analysis in financial news, with a focus on defining polarity based on stock price returns after the release of news articles. The authors report that their methodology, which utilizes a combination of recurrent neural network with long short-term memory units, shows improved forecasting accuracy when selecting news with the highest positive and negative scores as positive and negative news, respectively. They suggest several avenues for future research, including exploring different methods for defining polarity and using different deep learning methodologies. Overall, the findings of this study have significant implications for the field of financial forecasting and demonstrate the potential of deep learning methods in enhancing sentiment analysis [16].

The paper by Guo and Tuckfield (2020) investigates the effectiveness of news- based machine learning and deep learning methods in predicting stock indexes and individual stocks. The study compares the performance of machine learning and deep learning approaches using news data processed through natural language processing. The results suggest that deep learning outperforms ma- chine learning in predicting both stock indexes and individual stocks, with a 4.5 percentile improvement for stock indexes and a 3-percentile improvement for individual stocks. The authors discuss the implications of these findings and suggest future areas of research, such as using other types of data and optimizing the model for stronger generalization ability. This study highlights the potential of deep learning in predicting stock market trends and underscores the importance of data processing and model optimization in achieving accurate results [13].

FinBERT (Araci 2019) is used in this study to improve the accuracy of the text mining method. FinBERT is a pre-trained language model created by a Bloomberg research team. It is trained on a large corpus of financial documents, such as news articles, SEC filings, and earnings call transcripts, to improve its understanding of financial jargon, entity recognition, and sentiment analysis [25].

This study is unique in that it takes into account video news sentiment, which has largely been ignored in previous research. Furthermore, comparing previous models such as statistical methods and ML methods to see which method has the highest accuracy in real-world problems.

## METHODOLOGY AND DATA SET

Investors are constantly seeking new and innovative ways to forecast the future price of financial assets such as stocks and indices. The S&P500, a stock market index that tracks the performance of the 500 largest publicly traded companies in the United States, is of particular interest due to its impact on the broader economy.

We compare two popular methods of forecasting the future return rate of the S&P500: first of all, technical analysis, which includes statistical models such as Autoregressive Integrated Moving Average (ARIMA), Generalized Autoregressive Conditional Heteroskedasticity (GARCH), and also machine learning method, Long Short-Term Memory (LSTM). Second, text mining analysis, which utilizes the Bidirectional Encoder Representations from Transformers (BERT) model as well as FinBERT, a specialized version of BERT designed for financial text analysis.

Both technical analysis and text mining analysis use different techniques to analyze and predict the future return rate of the S&P500. Technical analysis employs statistical methods and machine learning models to analyze historical data, while text mining analysis uses natural language processing and machine learning techniques to analyze financial news.

## ARIMA

Auto Regressive Integrated Moving Average is a statistical model class that is used for time series analysis and forecasting. It is a widely used approach in the field of econometrics for modeling and predicting economic data such as stock prices, GDP, and inflation rates. ARIMA models assume that time series data can be described as a combination of autoregressive (AR), moving average (MA), and differencing terms. The basic mathematical equation for an ARIMA(p,d,q) model:

$$y(t) = c + \varphi(1)y(t-1) + \cdots + \varphi(p)y(t-p) - \theta(1)\varepsilon(t-1) - \cdots - \theta(q)\varepsilon(t-q) + \varepsilon(t) \quad (1)$$

where:
- y(t) is the value of the time series at time t
- c is a constant or intercept term
- (1) to (p) are the autoregressive (AR) coefficients of the model, representing the linear relationship between y(t) and its p previous values (y(t-1) to y(t-p)) - (1) to (q) are the moving average (MA) coefficients of the model, representing the linear relationship between the error term (t) and its q previous values ((t-1) to (t-q))
- (t) is the error term or white noise at time t
- d is the degree of differencing used to make the time series stationary, which is often necessary for ARIMA modeling.

**GARCH**

A class of statistical models known as Generalized Autoregressive Conditional Heteroskedasticity models is used to analyze and predict volatility in time series data. These models are widely used in finance, economics, and other fields where understanding and forecasting volatility are critical. The idea behind GARCH models is that the variance of a time series can change over time and that the variance is conditional on previous values of the series. GARCH models can model a series' time-varying volatility, which is important in many real-world time series such as stock prices, exchange rates, and commodity prices.

The basic GARCH model can be expressed mathematically as:

$$rt = \mu + \varepsilon t \quad (2)$$

$$\varepsilon t = \sigma t z t \quad (3)$$

$$\sigma^2{}_t = \alpha_0 + \alpha_1 \varepsilon^2{}_{t-1} + \beta_1 \sigma^2{}_{t-1} \ (4)$$

where $rt$ is the observed return at time t, μ is the mean of the return series, $\varepsilon t$ is the error term, $\sigma t$ is the conditional standard deviation at time t, and zt is a standardized error term with mean zero and variance one. The parameters α0, α1, and β1 are non-negative and reflect the persistence of volatility, the impact of past squared errors on current volatility, and the impact of past conditional variances on current volatility, respectively.

The GARCH model is estimated using maximum likelihood estimation, which involves finding the values of the model parameters that maximize the likelihood of observing the data.

**LSTM**

Long Short-Term Memory (LSTM) models are a type of recurrent neural net- work (RNN) that can analyze and predict sequential data with long-term dependencies. Hochreiter and Schmidhuber introduced LSTMs in 1997, and they have since become a popular tool for natural language processing, speech recognition, and time series analysis.

The ability of LSTMs to selectively remember and forget information over long periods of time makes them well-suited for modeling and predicting data sequences with complex patterns and dependencies. LSTMs accomplish this by employing memory cells, which allow them to keep a long-term memory of previous inputs and selectively update that memory based on new inputs. This makes LSTMs especially useful in applications like language translation, sentiment analysis, and stock price prediction, where understanding and modeling complex dependencies in sequential data is critical. The mathematical model for Long Short-Term Memory (LSTM) networks can be expressed as follows

$$ht = \sigma(Wxhxt + Whhht - 1 + bh)(5)$$

$$ct = f(Wxcxt + Whcht - 1 + bc)(6)$$

$$ot = \sigma(Wohxt + Wohht - 1 + bo)(7)$$

$$yt = g(V\,ht + c)\ (8)$$

where $ht$ represents the hidden state of the LSTM at time t, ct represents the cell state, xt represents the input at time t, and $yt$ represents the output at time t. The functions σ and g represent the sigmoid and activation functions, respectively, and f is the forget gate function. The LSTM model has three main components: the input gate, the forget gate, and the output gate. The input gate controls the flow of information into the cell state, while the forget gate selectively discards information that is no longer relevant. The output gate controls the flow of information from the cell state to the hidden state.

LSTM networks are trained using backpropagation through time (BPTT), which involves calculating the gradients of the loss function with respect to the model parameters at each time step and propagating them backwards through time. This allows the network to learn complex patterns and dependencies in sequential data, making LSTMs a powerful tool for a wide range of applications in natural language processing, speech recognition, and time series analysis.

**BERT**

Bidirectional Encoder Representations from Transformers is a deep learning model that has been pre-trained by Google AI Language. It has transformed natural language processing tasks by achieving cutting-edge performance on a variety of tasks such as text classification, question answering, and text generation. Unlike traditional language models, which process text sequentially, BERT is a transformer-based model that can consider both the preceding and following words to determine the context of a word or sentence. This enables BERT to capture complex linguistic phenomena such as word sense disambiguation and coreference resolution, resulting in significant accuracy and performance improvements.

The BERT model architecture is based on the transformer architecture, which is made up of a series of encoder and decoder layers that process input sequences in parallel, attention-based fashion. BERT pre-trains the model on a large corpus of text data with a bidirectional transformer encoder, then uses a masked language modeling objective to predict the missing words in a sentence. The pre-trained BERT model can be used as a starting point for a wide range of natural language processing tasks during fine-tuning, allowing it to achieve great performance with relatively little data.

BERT is a complex model that involves multiple layers of neural networks, but at its core, it uses a transformer-based architecture that can be expressed mathematically as follows: Given an input sequence X of tokens, BERT first applies an embedding layer to convert each token into a vector representation. The embedded tokens are then fed into a series of transformer encoder layers, which process the sequence in a parallel, attention-based manner. The output of the final transformer layer is a sequence of context-aware token representations, which can be used for downstream tasks such as text classification or question answering.

**FinBERT**

It is a BERT-based domain-specific language model designed specifically for analyzing financial text data. FinBERT has been pre-trained on a large corpus of financial text data, allowing it to comprehend the subtleties of financial language and terminology. It's been used for everything from sentiment analysis to financial document classification and financial forecasting. FinBERT has demonstrated cutting-edge performance on a variety of financial text datasets, making it an invaluable tool for financial analysts and researchers.

The FinBERT model architecture is similar to the BERT model architecture, but it has been fine-tuned on financial text data to achieve higher accuracy and performance. It encodes the input text data and extracts contextual representations using a deep transformer-based neural network. After that, the contextual representations can be used for a variety of financial text analysis tasks, such as sentiment analysis or document classification. FinBERT can significantly re- duce the amount of manual effort required for financial text analysis while also improving results accuracy.

**Data and evaluation metric**

In this study, we will use financial time series data (S&P500 price) from Invest- ing.com, it's a close daily price of the index. Text data from different financial video news (CNBC, Bloomberg, and yahoo fiNMnce) as our primary dataset. The data will consist of 2 years of financial video news and market data from Jan 2019 to Dec 2020, providing a comprehensive and robust understanding of market trends and behavior when we faced the pandemic [11].

The root mean square error (RMSE) is used to assess the accuracy of regression models. It is the square root of the average squared difference between the predicted and actual values in a dataset. The following is the formula for calculating RMSE:

$$RMSE = \sqrt{\frac{1}{n}\sum_{i=1}^{n}(\hat{y}_i - y_i)^2}$$

where
$\hat{y}_i$ is the predicted value,
$y_i$ is the actual value, and
n is the number of observations in the dataset.

## RESULTS AND DISCSSION

Six different models were used to forecast the price of the S&P500 in this study: ARIMA, GARCH, LSTM, BERT, and FinBERT. The dataset was divided into train and test sets to determine the optimal configuration for each model. The train set contained the first 75% of the data and the test set contained the remaining 25%. Different model parameters and settings were experimented with to optimize and find the configuration with the lowest RMSE on the test set. The RMSE of each model was compared to assess its forecasting performance [26-30].

### Technical analysis

In statistical models, the Python ”stats models” library was used to find the best ARIMA model for the return data using three commonly used criteria: Akaike Information Criterion (AIC), Bayesian Information Criterion (BIC), and Mean Squared Error (MSE). After comparing the models using these criteria, it was discovered that the best ARIMA model for the data had p=4, q=1, and d=0.

This implies that the best fit to the data is provided by a model with a moderate amount of autoregressive and moving average terms but no differencing term. The chosen model can be used for forecasting and other return data analysis. Using the test data, the selected ARIMA model with p=4, q=1, and d=0 was further evaluated by calculating its root mean squared error (RMSE). The final RMSE was found to be 0.0103, indicating that the model fits the data well and can provide accurate forecasts.

In GARCH model selection, the Python ”arch” library makes it simple to estimate and choose the best GARCH model. The library includes functions for fitting GARCH models, parameter estimation, and forecast generation. It also has functions for calculating the AIC, BIC, and HQIC criteria, making it easier to choose the best model based on these criteria.

When the RMSE values of the GARCH(1,1) and ARIMA(4,0,1) models were compared, the former had a higher RMSE value of 0.5922, indicating that the ARIMA model may be better suited for this dataset. This could be due to the highly stationary(Figure 2) nature of the dataset, which may have limited the GARCH model’s ability to capture volatility patterns. Finally, for highly stationary return data, an ARIMA model may outperform a GARCH model in terms of forecasting accuracy.

In machine learning models, an LSTM model is used to forecast time series data. The Keras library is used to create and train the model, with the input layer consisting of two LSTM layers followed by a Dense output layer. After that, the model is compiled with the ’adam’ optimizer and the ‘mean squared error’ loss function. Thus, the code computes the root mean squared error (RMSE) to assess the model’s performance on the test set, with an RMSE of 0.5646 indicating promising forecasting accuracy.

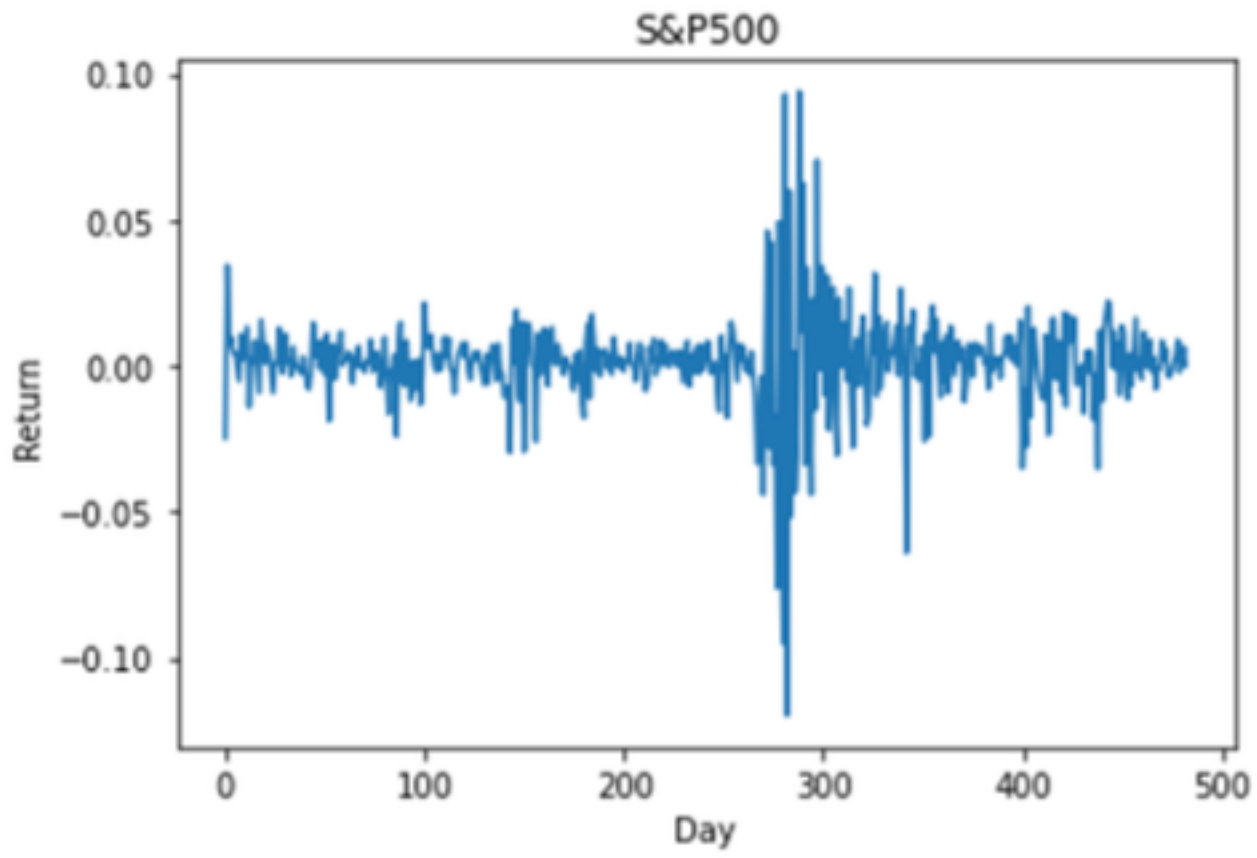

Figure 2: Daily return rate

The code trains an LSTM model using the training data and saves the training history. The history is then used to plot the training and validation loss function over the course of 100 epochs using matplotlib (Figure 3).The plot dis- plays the loss values on the y-axis and the number of epochs on the x-axis, with separate lines for the training and validation loss. The legend on the plot identifies the two lines as the train and test loss, respectively. The plot provides a visual representation of the model's performance during training and can be used to determine if the model is overfitting or underfitting the data. If the loss for the validation set starts to increase while the training set loss continues to decrease, this may indicate that the model is overfitting to the training data and not generalizing well to new data.

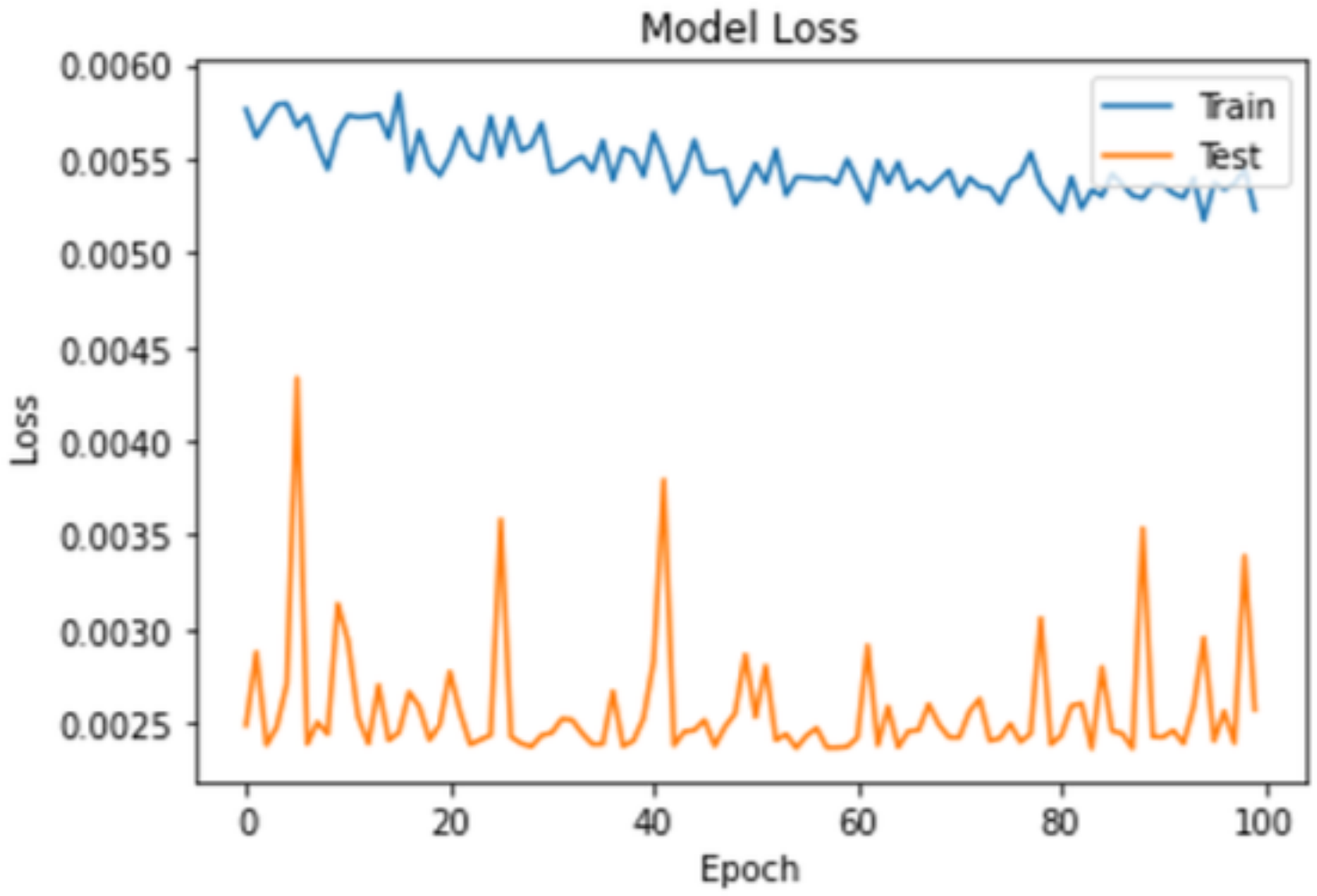

Figure 3: LSTM model

## Text mining analysis

The BERT transformer is utilized to create a deep learning model for regression tasks. The AutoConfig and BertRegresser functions from the Hugging Face Transformers library are utilized to configure and initialize the model, respectively. The Adam optimizer is utilized with a learning rate of LR to optimize the model's parameters, while the mean squared error loss function is used to compute the loss between the predicted and target outputs. An RMSE value of 0.0120 for a BERT model indicates that the model has achieved a high level of accuracy in predicting the target variable. However, the FinBERT model, which has been pre-trained on a large corpus of financial text, can effectively learn the underlying patterns in financial data and make accurate predictions. We fine-tuned the FinBERT model on a dataset of financial news articles and achieved an RMSE value of 0.010284, which is a great result for a regression task. The effectiveness of the FinBERT model versus a BERT model may differ depending on the dataset and the amount of training data available. FinBERT may outperform BERT if the dataset is relatively small because it has been specifically trained on financial texts and thus may better capture the domain- specific language and nuances of the financial domain.As a result, when choosing a pre-trained language model for a particular task, it is critical to consider the nature of the task, the characteristics of the dataset, as well as the strengths and limitations of the available models. To determine which model performs best, it may be necessary to experiment with different models and fine-tune them on the specific task.

## CONCLUSION AND FUTURE WORK

The goal of this study was to forecast the return of the S&P500 using text mining and technical analysis methods. We used the dataset from investing.com and various financial platforms that included two years of data. Using techniques such as word embedding, we converted text data into numerical vectors, which we then fed into regression models. Technical analysis, on the other hand, involved the use of traditional methods such as ARIMA, GARCH, and some machine learning approaches such as the LSTM model. The accuracy of the predictions made by the various methods of forecasting the S&P500's future return rate was evaluated using only one metric: Root Mean Squared Error (RMSE).

We discovered that the FinBERT model outperformed the ARIMA and LSTM models in predicting the S&P500 price because it had a lower RMSE value. This suggested that the FinBERT model, rather than the ARIMA or LSTM models, could be a more reliable tool for forecasting the S&P500 price. AccordIng to the data, financial news appeared to be more useful than prices during the COVID-19 pandemic.

Finally, we believe that the FinBERT model has the potential to transform the field of financial analysis and prediction, resulting in new insights and discoveries. However, more research is required to fully exploit the FinBERT model's potential and assess its effectiveness across various financial datasets.